# Study Of Si-Ge Interdiffusion With a High Phosphorus Doping Concentration

Feiyang Cai[1], Dalaver H. Anjum[2], Xixiang Zhang[3] and Guangrui (Maggie) Xia[1]

[1]Department of Materials Engineering, University of British Columbia, Vancouver, BC V6T1Z4, Canada, Tel: +1-604-8220478, Fax: +1-604-8223916, E-mail: gxia@mail.ubc.ca

[2]Imaging and Characterization Core Lab, King Abdullah University of Science and Technology, 23955-6900 Thuwal, Saudi Arabia

[3] Materials Science and Engineering, King Abdullah University of Science and Technology, Thuwal 23955 - 6900, Saudi Arabia

1. Introduction

Germanium is the most silicon compatible semiconductor in terms of lattice structure, solubility, alloying, growth and processing compatibility. Germanium on silicon (Ge/Si) structures are widely used in semiconductor devices such as Ge-on Si lasers [1-3], infrared photodetectors [4-6] and chemical sensors [7].

For Ge lasers, adding high concentrations of n-type dopants, normally phosphorus (P), in Ge is crucial to occupy the electron energy states in the indirect conduction valley at L point [8-10]. Many efforts have been made on that. Delta-doped layer and gas immersion laser doping were used to achieve up to $5 \times 10^{19}$ cm$^{-3}$ activation of P doping [10-11]. Spin-on dopant process and multiple implantation were also successful in doping Ge up to $1 \times 10^{20}$ cm$^{-3}$ [12-13].

However, high n-type doping levels generate two side effects: 1) larger optical loss due to free carrier absorption and 2) much faster Si-Ge interdiffusion that changes the Ge/Si interfaces into SiGe alloy transition regions, which delays lasing [1,2,14]. This intermixing layer can be polished off after the original Ge/Si wafer is flip bonded to a handle wafer, and thinned down to form a Ge-on-insulator (GOI) structure. However, silicon dioxide has a much worse thermal conductivity than silicon, and Ge on Si substrates are better than GOI structures to dissipate heat generated by the Ge lasers. For Ge-on-Si structures, the as-grown dislocation density is normally in the $10^9$ to $10^{10}$ cm$^{-2}$ range, too high for device applications. A defect annealing step is commonly used in the epitaxial reactor following the growth to reduce the dislocation density to $10^6$ to $10^7$ cm$^{-2}$ range.

P enhanced interdiffusion has been reported for SiGe alloys with a low Ge molar faction ($x_{Ge}$) of 26.5%, which was attributed to the indirect interaction between Ge and point defects released by P-defect clusters [15]. Another work by H. Takeuchi and P. Ranade, reported that n-type doping by arsenic also enhanced Ge and Si interdiffusion [16, 17].

Our recent study showed that high P doping greatly accelerates Si-Ge interdiffusion due to the Fermi-level effect [18]. Si-Ge interdiffusion can change a Ge layer into a SiGe alloy, which delays the lasing of Ge lasers. Previous studies on Si-Ge interdiffusion with doping are mainly phenomenological observations using Ge-on-Si structures [9-11, 18], and no quantitative modeling is available for Si-Ge interdiffusion with high P doping.

According to reference [19, 20], the diffusivity of P in SiGe material systems is a function of both the Ge fraction ($x_{Ge}$) and the concentration of P ($C_P$). In the meantime, the interdiffusivity of Si-Ge is also strongly dependent on $x_{Ge}$ and $C_P$ [18, 21]. On top of that, P segregates to regions with lower Ge content [18]. That means P diffusion, segregation, and Si-Ge interdiffusion are coupled, and it is very hard to extract P diffusivity and Si-Ge interdiffusivity from experimental data especially in a large Ge molar fraction range as all three processes occur simultaneously.

On top of the complicated and coupled phenomena (Si-Ge interdiffusion, P diffusion and segregation), experimentally, P forms a very high segregation peak at Ge/Si interfaces larger than $10^{20}$ cm$^{-3}$ [14]. This peak is very hard to measure accurately by SIMS as SIMS accuracy is greatly degraded due to the knock-on effect at interfaces and the mixing effect of narrow concentration peaks [22]. Another difficulty is the existence of Ge seeding layers that is required for the subsequent Ge layer growth, where high concentrations of defects exit. These defects play an important role in the mass transport of P, Si, and Ge, which add another layer of complexity in the modeling of Si-Ge interdiffusion, P diffusion and segregation. Therefore, Ge-on-Si structures for Ge lasers are not ideal for Si-Ge interdiffusion studies.

In this work, new structures were designed and measured to circumvent this problem. A quantitative model of the Fermi-level effect on Si-Ge interdiffusion was proposed to model the extrinsic interdiffusion behavior. According to our knowledge, the quantitative modeling of Si-Ge extrinsic interdiffusion has not been reported previously.

## 2. Experiments
### 2.1 Structure Design, Growth, and Annealing

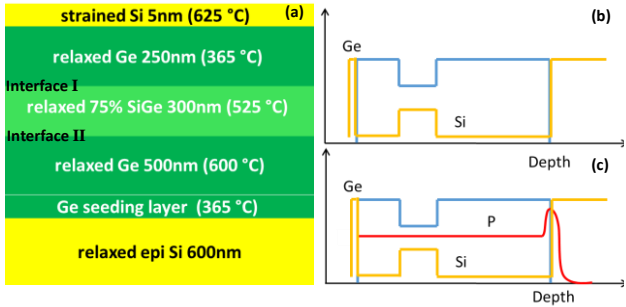

**Figure 1** Schematic diagrams of the samples used in this work: (a) sample structure and growth temperature (b) depth profile of sample 7450 with no P doping and (c) depth profile of sample 7452 with P doping concentration at around $5\times10^{18}$ cm$^{-3}$.

To avoid the complications from Ge seeding layers, a Ge/Si$_{1-x}$Ge$_x$/Ge multi-layer structure on Si substrates with x = 75% was designed for interdiffusion study (Figure 1). The Si-Ge interdiffusion region of interest was then moved to the top Ge/Si$_{1-x}$Ge$_x$ (denoted as the Interface I) and Si$_{1-x}$Ge$_x$/bottom Ge (denoted as the Interface II) interfaces, which was 0.5 micron away from the Ge seeding layer/Si interfaces. The 25% Ge molar fraction change at interfaces was chosen to avoid too much P segregation and P diffusivity change due to large Ge fraction changes such as those at Ge/Si interfaces. Two P doping configurations were designed: Sample 7450 (S7450) with no P doping and Sample 7452 (S7452) with P doping at mid-$10^{18}$ cm$^{-3}$ in the Ge/Si$_{1-x}$Ge$_x$/Ge multi-layer structure.

Both samples were grown on 6 inch Czochrolski (CZ) (100) Si wafers in an Applied Materials "Epi Centura" system. For S7450, a thin Ge seeding layer was grown at 365 °C. Next, a 500 nm Ge layer was deposited at 650 °C, and then a 300 nm of Si$_{0.25}$Ge$_{0.75}$ layer was grown at 525 °C. On top of the Si$_{0.25}$Ge$_{0.75}$ layer, another 250 nm Ge layer was grown at 365 °C under the seeding layer growth condition. Finally, a 5 nm thin silicon cap was grown on top at 625 °C to prevent Ge evaporation during annealing. For S7452, it was grown by the same procedure except that P was in-situ doped with mid-$10^{18}$ cm$^{-3}$ concentration level during the growth of the Ge/SiGe/Ge structure.

Before annealing, samples were capped with a SiO$_2$ layer at 80 °C by plasma-enhanced chemical vapor deposition (PECVD) to prevent Ge outdiffusion. Inert annealing was performed in nitrogen atmosphere using an enclosed Linkam TS1200 high-temperature heating stage. Two annealing conditions were used for both samples: 1) 750 °C for 120 min; 2) 800 °C for 30 min. Trial annealing was performed and then final annealing conditions were chosen such that Si-Ge interdiffusion is not too little to be detected by secondary ion mass spectrometry (SIMS) or too large for the diffusivity extraction method, Boltzmann-Matano analysis, to be not applicable, i.e., the top and bottom Ge concentration should stay close to 100% such that we can treat interdiffusion at Interface I and II as the interdiffusion of a diffusion couple.

The temperature ramp up rate was set to be 100 °C/min, and the cooling was by water cooling, whose rate was about 200 °C/min. The amount of diffusion during the temperature ramp up and ramp down was negligible compared to the diffusion in the isothermal annealing step, which was later confirmed by simulations using the model established as described in 3.1.

SIMS measurements were employed to obtain the as-grown and annealed Ge and P profiles of S7450 and S7452. High resolution X-ray diffraction (HRXRD) measurements were performed to get the strain status in both samples. Transmission electron microscope (TEM) was used to check the material quality of the as-grown and annealed samples.

### 2.2 Strain and Interdiffusion Characterizations by XRD

In order to study the strain status in the Ge and SiGe layer, HRXRD measurements were performed using a PANalytical X'Pert PRO MRD with a triple axis configuration. Figure 2 shows the (004) symmetric XRD scans of as-grown and annealed samples. For each sample, the strongest peak is from the Si substrate, and the middle peak is from the SiGe layer. The second strongest peak (the right peak) is from the Ge layers. For S7450 and S7452, the peak positions are very close, which indicates the similar strain status for both samples. Compared to the as-grown samples, the Ge and SiGe peaks of annealed samples are asymmetric and broadened on the side towards each other. This broadening is attributed to Si-Ge interdiffusion. For annealed S7452, the broadening is much more obvious than annealed S7450, indicating larger Si-Ge interdiffusion with P doping. With the Si substrate peak as a reference, we can see that the peak from the SiGe layer shifted towards left after annealing. This result suggests that after annealing, the SiGe layer has a lower strain due to in-plane strain relaxation.

In addition, based on the peak positions from the (004) symmetric XRD scans and Ge fractions from the SIMS data in Section 2.3, the strain levels in the Ge and the SiGe layers were extracted using the PANalytical Epitaxy software package, which are summarized in Table 1. For S7450 and S7452, the strain status for corresponding layers is very close. The difference between strains is within 0.05%, which suggests P doping did not affect the strain status of samples significantly. Besides, the Ge and SiGe layers are both tensily strained. According to Xia et al.'s observation, tensile strain up to 1% does not affect Si-Ge interdiffusion [23]. Therefore, there was no need to consider tensile strain in the interdiffusion modeling.

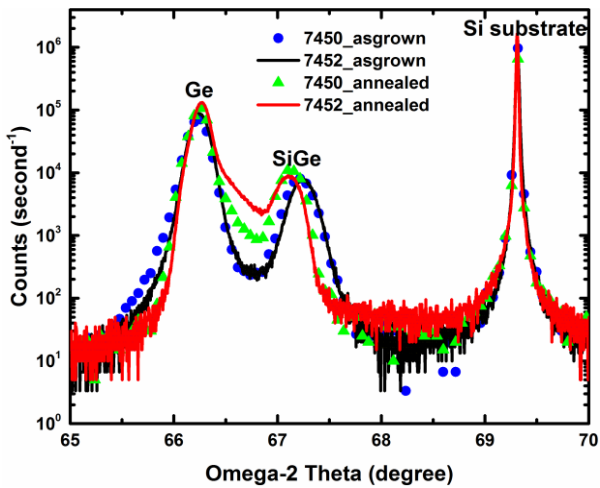

**Figure 2** (004) symmetric XRD scans of the four structures.

| Layer | Strain | | | |
|---|---|---|---|---|
| | S7450 As-grown | S7450 Annealed | S7452 As-grown | S7452 Annealed |
| Ge | 0.10% | 0.16% | 0.13% | 0.16% |
| SiGe | 0.52% | 0.38% | 0.50% | 0.33% |

**Table 1** Strain statuses of the Ge and SiGe layers in sample 7450 and sample 7452 calculated from the XRD measurements. The corresponding annealing condition is 750 °C 120 minutes.

2.3 Concentration Profiling by SIMS

The Ge and P concentration vs. depth profiles were obtained by SIMS. The samples were sputtered with a 1 keV Cesium (Cs) ion beam which was obliquely incident on the samples at 60° off the sample surface normal. The sputter rate was calibrated by a stylus profilometer that measured the sputtered carter depth. With the known sputter rate variation with SiGe composition, the sputter rate was corrected on a point-by-point basis. The measurement uncertainty in Ge molar fraction is ± 1%. The depth uncertainty is about 5%.

Figure 3 shows the Ge concentration vs. depth profiles obtained by SIMS for S7450 and S7452. Due to the non-uniformity of the epitaxial growth and errors in the SIMS depth calibration, which can add up and result in 5-10% error, the as-grown and annealed SIMS profiles are not aligned precisely. To compare the amount of interdiffusion, the Ge profiles are shifted laterally to offset the non-uniformity and depth errors such that the difference between two samples can be easily seen and compared.

We can see that S7450 and S7452 have similar as-grown profiles and they are very steep at the interfaces. After annealing, the Ge molar fractions at the left and right end of the Ge/SiGe/Ge sandwich structure are still at 100%, and that in the middle of the SiGe valley stay flat at 77%, which means that the interdiffusion at Interface I and II are independent from each other and they can be treated as interdiffusion from two independent diffusion couples.

For annealed S7450 (undoped), the interdiffusion of Si-Ge show a large Ge fraction dependence, where much more diffusion happens in higher Ge regions than in lower Ge regions. For annealed S7452 (P-doped), however, the difference is not as obvious. By comparison, in $x_{Ge}$ > 0.9 region, the amount of interdiffusion in S7452 is close to that in S7450. However, in $x_{Ge}$ < 0.9 region, S7452 has significantly more interdiffusion, which also means that interdiffusion with P doping is much less Ge concentration dependent.

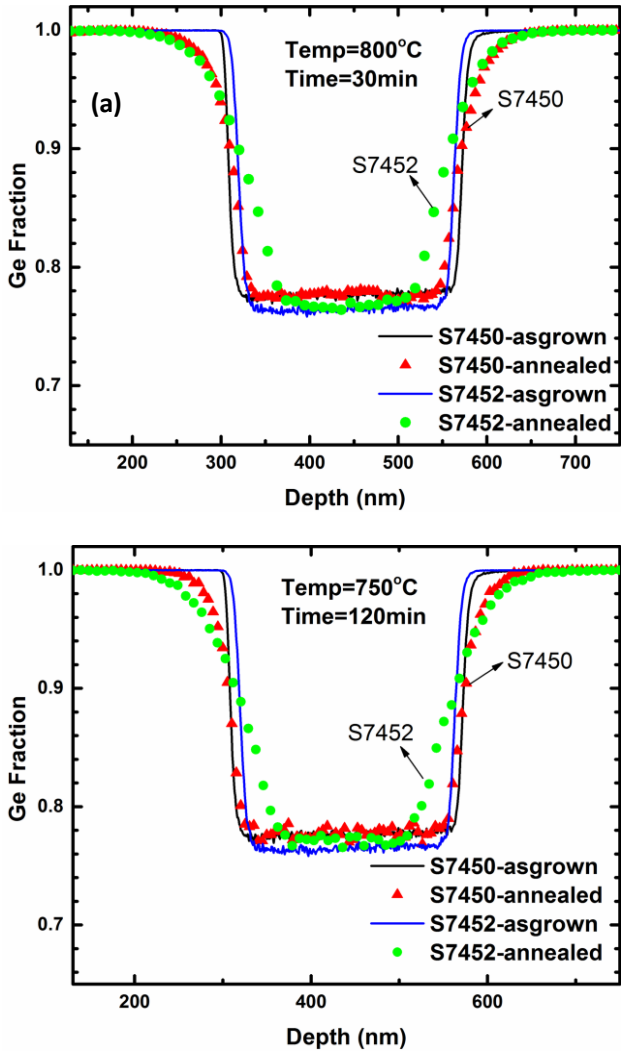

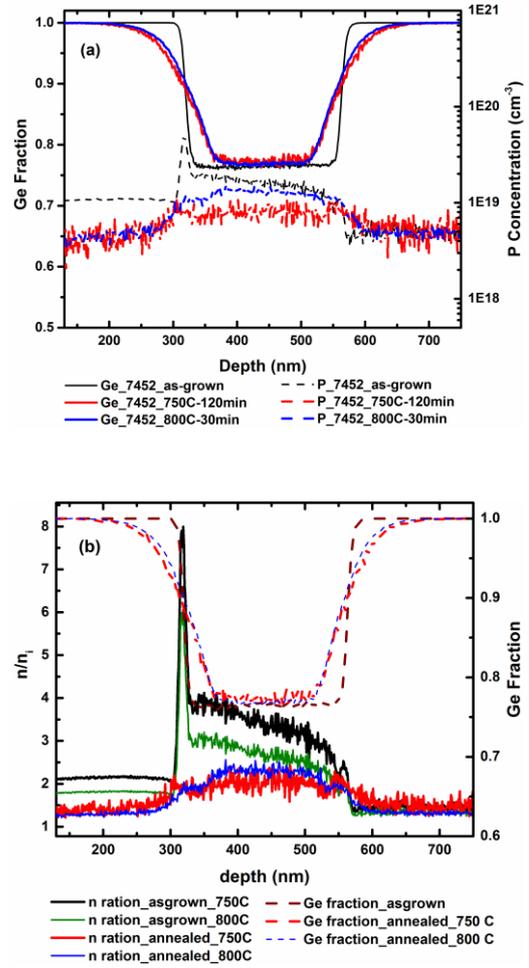

Figure 3 Ge profiles measured by SIMS. (a) samples annealed at 750 °C, 120 minutes; (b) samples annealed at 800 °C, 30 minutes.

Figure 4 (a) Ge and P profiles in S7452 showing the P concentration variation along with Ge fraction. The Ge and P profiles are shifted laterally for clarity. (b) The ratio of $n/n_i$ profile of S7452 at 750 °C and 800 °C, before and after annealing. Dash lines are the Ge fraction profiles of S7452.

According to semiconductor diffusion theories, charged defect concentrations depend on the Fermi level and the ratio of the electron density over the intrinsic electron density $n/n_i$ [22]. $n/n_i$ is an important parameter for the Fermi-level effect calculations. Only when $\frac{n}{n_i} > 1$, Fermi-level effect needs to be considered and higher $n/n_i$ means stronger doping introduced diffusivity dependence. This also applies to the case of interdiffusion, which is mediated by point defects as well. Figure 4 shows the P profiles and the ratio of $n/n_i$ in S7452. The calculations of $n(x_{Ge})$ and $n_i(x_{Ge})$ were described in 3.1.

As expected from the design, in Figure 4 (a), P segregation is much more reduced compared to the cases of P at Ge seeding layer/Si interfaces. It is still observable as there are higher P concentrations in the SiGe layer than in the surrounding Ge layers. In as-grown S7452, a P peak appears at the Interface I. This P peak was formed during growth, and the P concentration in the top Ge layer is about two times of that in the bottom Ge layer, which is due to the growth temperature change. After annealing, P concentration decreased and the P peak at the Interface I also disappeared. In Figure 4 (b), we can see the ratios of $n/n_i$ are all larger than 1 before and after annealing, and the ratio in the lower Ge regions ($x_{Ge}$ < 0.9) is higher. This is consistent with the observation that more Si-Ge interdiffusion happened in the lower Ge region of S7452 with more P doping. For higher Ge regions, however, the ratios of $n/n_i$ are close to one which explains the much closer Ge profiles in the higher Ge regions ($x_{Ge}$ > 0.9) of S7452 and S7450.

2.4 TEM and Defect Density Characterization

TEM images were collected to observe the interface changes and estimate the threading dislocation density

(TDD) with a Titan3 80-300 TEM. The thickness of the TEM-specimens was estimated to be around 100 nm by electron energy loss spectroscopy (EELS). The TDs are estimated directly from these TEM images of the cross section of the samples. To obtain the TDD from cross-sectional TEM images, the number of TDs in a horizontal line were counted first. The dislocation density per unit length is the number of dislocations divided by the length in the TEM image. Finally, the TDD is the square of the dislocation density per unit length [24].

Figure 5 shows the cross-sectional TEM images of S7450 and S7452 before and after annealing. We can see that for each sample, before annealing, there were clear boundaries at the Interface I and Interface II. However, after annealing, those boundaries became disruptive and wavy. Compared with annealed S7450, the boundaries of S7452 are more disruptive. Massive misfit dislocations were generated on the boundaries during annealing to relieve the tensile strain between layers, as shown in XRD results.

In Figure 5, TDs were pointed out by white arrows. Table 2 lists the average TDDs for S7450 and S7452 measured by TEM before and after annealing. The TDDs are similar for upper and lower interface. It is found that under same conditions, the ratio between TDD of S7450 and S7452 is in the range of (0.5-2), which suggests P doping did not affect the TDD of the sample significantly.

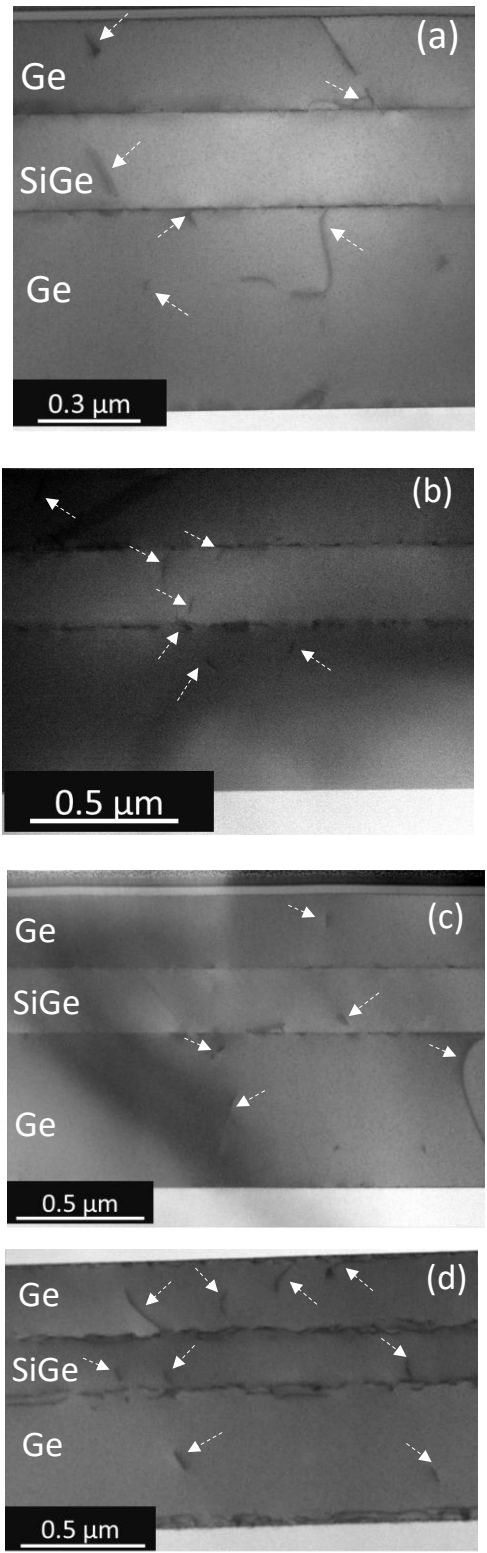

Figure 5 Cross-sectional (bright field) TEM of S7450 (no P) and S7452 (with P). (a) as-grown S7450; (b) annealed S7450; (c) as-grown S7452; (d) annealed S7452. Threading dislocations are pointed out by white arrows. For annealed samples, the thermal budget is at 800 °C for 30 minutes.

| Sample ID | Ge/SiGe Interface | SiGe/Ge Interface |
|---|---|---|
| S7450 as-grown | $4.3 \times 10^8$ cm$^{-2}$ | $1.2 \times 10^9$ cm$^{-2}$ |
| S7452 as-grown | $2.7 \times 10^8$ cm$^{-2}$ | $5.4 \times 10^8$ cm$^{-2}$ |
| S7450-annealed | $6.2 \times 10^8$ cm$^{-2}$ | $6.5 \times 10^8$ cm$^{-2}$ |
| S7452-annealed | $1.1 \times 10^9$ cm$^{-2}$ | $7.1 \times 10^8$ cm$^{-2}$ |

**Table 2** Average TDD values of Ge/SiGe (upper) interface and SiGe/Ge (lower) interface of S7450 and S7452. The thermal budget for annealed samples is at 800°C for 30 minutes.

# 3 Interdiffusivity Modeling and Calculations

## 3.1 Review of related semiconductor diffusion theories

It has been widely accepted that diffusion and interdiffusion in crystalline semiconductors are mediated by point defects, which can be in charged states [22]. The (inter)diffusivity dependence on electron concentration n in an n-type doped semiconductor can be expressed as:

$$D = D_0 + D_-\left(\frac{n}{n_i}\right) + D_=\left(\frac{n}{n_i}\right)^2 + D_+\left(\frac{n}{n_i}\right)^{-1} + D_{++}\left(\frac{n}{n_i}\right)^{-2} \text{ [22]}, \quad (1)$$

where parameter $D_0$, $D_-$, $D_=$, $D_+$, $D_{++}$ are the (inter)diffusivity associated with neutral, single negatively charged, double negatively charged, single positively charged, and double positively charged defects respectively. D is the total (inter)diffusivity. Similarly, for p-type doped cases,

$$D = D_0 + D_-\left(\frac{p}{n_i}\right)^{-1} + D_=\left(\frac{p}{n_i}\right)^{-2} + D_+\left(\frac{p}{n_i}\right) + D_{++}\left(\frac{p}{n_i}\right)^2. \quad (2)$$

The (inter)diffusivity under intrinsic conditions, i.e. the doping level is much less than $n_i$, is given by:

$$D(n_i) = D_0 + D_- + D_= + D_+ + D_{++}. \quad (3)$$

Experimentally, for an n-type doped case as in (1), if the total (inter)diffusivity D is linearly dependent on the ratio $\frac{n}{n_i}$, that means $D_-\left(\frac{n}{n_i}\right)$ term is the dominant term, and the diffusion is mostly mediated by single negatively charged point defects. If the total (inter)diffusivity, D, is quadratically dependent on the ratio $\frac{n}{n_i}$, then $D_=\left(\frac{n}{n_i}\right)^2$ term is dominant, and the (inter)diffusion is mostly mediated by double negatively charged point defects.

For n-type doped semiconductors, dopant diffusion is mainly through neutral and negative charged point defects. The fourth term $D_+$, and the fifth term $D_{++}$ are neglected. For Si-Ge interdiffusion, according to the observation of Gavelle *et. al.* where the Ge layer is boron doped [25], Si-Ge interdiffusion is retarded, which suggests that the contributions from positively charged defects are negligible. Therefore, only negatively charged defects are considered in the modeling and we can approximate the intrinsic and the extrinsic interdiffusivity respectively as

$$\widetilde{D} \approx \widetilde{D}_0 + \widetilde{D}_- + \widetilde{D}_=, \quad (4)$$

and

$$\widetilde{D} \approx \widetilde{D}_0 + \widetilde{D}_-\left(\frac{n}{n_i}\right) + \widetilde{D}_=\left(\frac{n}{n_i}\right)^2 \quad (5)$$

Considering the charge neutrality equation $n = p + C_P$ and $n_i^2 = np$, the electron concentration n of the P-doped Si$_{1-x}$Ge$_x$ samples can be expressed as:

$$n(x_{Ge}) = \frac{C_P + \sqrt{C_P^2 + 4n_i^2(x_{Ge})}}{2}. \quad (6)$$

$n_i(x_{Ge})$ was calculated as $n_i(x_{Ge}) = n_i(x_{Ge})\exp(\frac{\Delta E_g(x_{Ge})}{kT})$ for $x_{Ge}$ < 0.80, and $n_i(x_{Ge}) = n_i(x_{Ge} = 0.8) * \frac{1-x_{Ge}}{0.2} + n_i(x_{Ge} = 1) * \frac{x_{Ge}-0.8}{0.2}$ for $x_{Ge}$ > 0.80.

## 3.2 Modeling of the Fermi-level Effect on Si-Ge Interdiffusion

To model $\widetilde{D}$, we need to first determine the participating defects. In Si$_{1-x}$Ge$_x$, it has been shown that both Si and Ge diffusion is vacancy-mediated when Ge fraction is larger than 0.3 [26-27]. In sample 7452, the Ge fraction is above 0.75. Therefore, the total interdiffusion coefficient is the sum of the individual contributions from neutral, single negatively charged and double negatively charged vacancies:

$$\widetilde{D} = \sum_{r=0}^{2}\widetilde{D}_{V^{r-}} = \frac{1}{C_0}\sum_{r=0}^{2}f_r C_{V^{r-}}^{eq}\widetilde{D}_{V^{r-}} \text{ [20]}, \quad (7)$$

where $C_0$ is the atom density of Si$_{1-x}$Ge$_x$ with $x_{Ge}$ ranges from 0.75 to 1. In this range, the atomic density of $C_0$ only varies by about 3%, so we can approximate that $C_0$ is independent of x. $f_r$ is the diffusion correlation factor. For diffusion via vacancies in diamond structures like in Si$_{1-x}$Ge$_x$, $f_r$ is considered to be independent of the charge state and is set to be $f_r = f_V = 0.5$ [28]. $C_{V^{r-}}^{eq}$ is the thermal equilibrium concentration of $V^{r-}$ point defects, and $\widetilde{D}_{V^{r-}}$ is the interdiffusivity mediated by $V^{r-}$ with $r \in \{0,1,2\}$, respectively. Assume $\widetilde{D}_{V^-} = m_1\widetilde{D}_{V^0}$ and $\widetilde{D}_{V^{2-}} = m_2\widetilde{D}_{V^0}$, where $m_1, m_2$ are fitting parameters. Therefore, Equation (7) can be transformed as:

$$\widetilde{D}(n) = \widetilde{D}_{V^0}\frac{C_{V^0}^{eq}}{2C_0}\left(1 + m_1\frac{C_{V^-}^{eq}}{C_{V^0}^{eq}} + m_2\frac{C_{V^{2-}}^{eq}}{C_{V^0}^{eq}}\right). \quad (8)$$

Charged point defects have energy levels in the bandgap, and the occupation of defect related energy levels depends on the position of the Fermi level $E_f$, which is a function of the dopant concentration. If the doping concentration exceeds the intrinsic carrier concentration n$_i$, the Fermi level $E_f$ will deviate from its intrinsic position $E_i$. The ratio of the charged vacancy concentration to the neutral vacancy concentration is given by Equation (9) and (10) in [22].

$$C_{V^-}^{eq} = C_{V^0}^{eq} exp\left(\frac{E_F - E_{V^-}}{kT}\right) \quad (9)$$

$$C_{V^{2-}}^{eq} = C_{V^0}^{eq} exp\left(\frac{2E_F - E_{V^-} - E_{V^{2-}}}{kT}\right) \quad (10)$$

Combining Equation (8-10), $\widetilde{D}(n)$ can be expressed as:

$$D(n) = \widetilde{D}_{V^0} \frac{C_{V^0}^{eq}}{2C_0}\left[1 + \sum_{r=1}^{2}\left(\frac{n}{n_i}\right)^r m_r exp\left(\frac{rE_i - \sum_{n=1}^{r} E_{V^{n-}}}{kT}\right)\right].$$
(11)

When $n = n_i$, the total interdiffusion coefficient is equal to the intrinsic interdiffusivity, $\widetilde{D}(n_i)$. Therefore, the ratio between extrinsic and intrinsic diffusion coefficient is:

$$\frac{\widetilde{D}(n)}{\widetilde{D}(n_i)} = \frac{1 + m_1 exp\left(\frac{E_i - E_{V^-}}{kT}\right)\left(\frac{n}{n_i}\right) + m_2 exp\left(\frac{2E_i - E_{V^-} - E_{V^{2-}}}{kT}\right)\left(\frac{n}{n_i}\right)^2}{1 + m_1 exp\left(\frac{E_i - E_{V^-}}{kT}\right) + m_2 exp\left(\frac{2E_i - E_{V^-} - E_{V^{2-}}}{kT}\right)},$$
(12)

For simplicity, we denote:

$$\beta = m_1 exp\left(\frac{E_i - E_{V^-}}{kT}\right), \quad (13)$$

$$\gamma = m_2 exp\left(\frac{2E_i - E_{V^-} - E_{V^{2-}}}{kT}\right), \quad (14)$$

where $\beta$ is related to the interdiffusion process mediated by V$^-$ point defects and $\gamma$ is related to the interdiffusion process mediated by V$^{2-}$ point defects. Therefore the interdiffusion coefficient as the function of electron concentration can be described as:

$$\frac{\widetilde{D}(n)}{\widetilde{D}(n_i)} = \frac{1 + \beta \frac{n}{n_i} + \gamma \left(\frac{n}{n_i}\right)^2}{1 + \beta + \gamma} \equiv FF, \quad (15)$$

where term $\frac{1 + \beta \frac{n}{n_i} + \gamma \left(\frac{n}{n_i}\right)^2}{1 + \beta + \gamma}$ is referred as the Fermi-enhancement factor (FF) in the discussions below.

For $\beta$ and $\gamma$, the energy term $rE_i - \sum_{n=1}^{r} E_{V^{n-}}$ ($r \in \{1, 2\}$) is a function of Ge fraction in Si$_x$Ge$_{1-x}$. Due to limited literature resources of the energy levels of $V^-$ and $V^{2-}$ in SiGe, i.e. $E_{V^-}(x_{Ge})$, and $E_{V^{2-}}(x_{Ge})$, for $0.75 < x_{Ge} < 1$, these terms were linearly interpolated between the values in Si and Ge, i.e.:

$$rE_i - \sum_{n=1}^{r} E_{V^{n-}} \equiv A_{r,SiGe}(x_{Ge}) = A_{r,Si}(1 - x_{Ge}) + A_{r,Ge} x_{Ge} \quad for \ (r \in \{1, 2\}). \quad (16)$$

For Si, $A_{1,Si} = 0.1383 \ eV$ and $A_{2,Si} = -0.1835 \ eV$ [22]. For Ge, $A_{1,Ge} = -0.1134 \ eV$ and $A_{2,Ge} = -0.0866 \ eV$ [20].

With these parameters, the exponential terms in Equation (13) and (14) can be calculated. For $0.75 < x_{Ge} < 1$ at 750 to 800 °C, $exp\left(\frac{E_i - E_{V^-}}{kT}\right)$ is in the range of 0.58-0.28, and $exp\left(\frac{2E_i - E_{V^{1-}} - E_{V^{2-}}}{kT}\right)$ in the range of 0.28-0.39. The only unknown parameters in this model are $m_1$ and $m_2$.

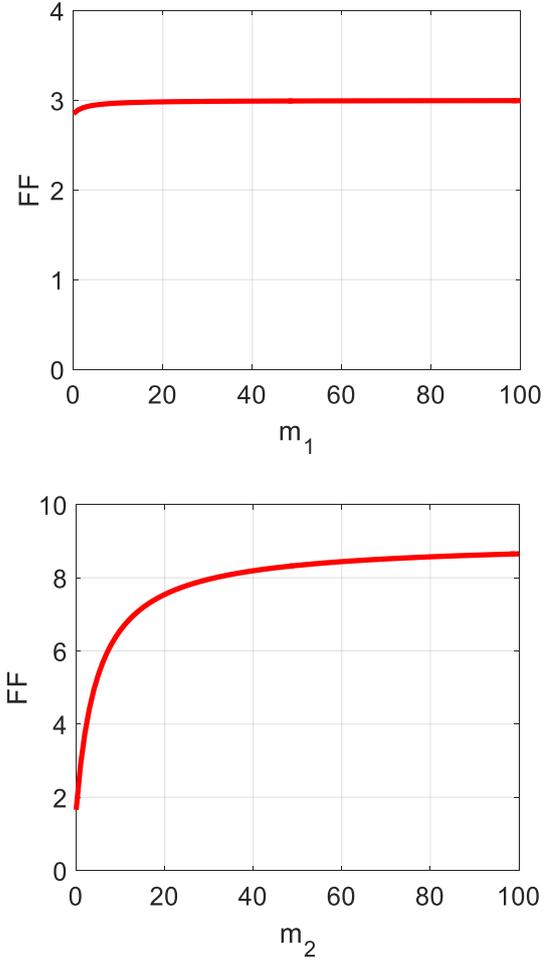

Figure 6 (a) Impact of $m_1$ to FF with $m_2 = 1$; (b) Impact of $m_2$ to FF with $m_1 = 1$. For both calculations, the temperature was fixed at 750 °C, $x_{Ge} = 0.80$ and $n/n_i = 3$.

Examples of the impacts of $m_1$ and $m_2$ to FF are shown in Figure 6. When all other parameters remain unchanged, as $m_1$ increases, the value of FF gradually approaches $n/n_i$, which suggests that the interdiffusivity is almost proportional to $n/n_i$. According to Equation (12), this indicates that the interdiffusion is dominated by $V^-$ point defects. Similarly, as $m_2$ increases, the value of FF gradually approaches $n^2/n_i^2$, which suggests that interdiffusion is dominated by $V^{2-}$ point defects. The values of $m_1$ and $m_2$ were extracted by fitting to the SIMS profiles, which is discussed in 4.1.

3.3 Intrinsic Interdiffusivity Extraction

In Equation (15), $\widetilde{D}(n_i)$ is the extracted interdiffusivity of the undoped sample S7450. From the above discussions, S7450 and S7452 have similar strain and defect levels, and similar Ge profiles before annealing. Therefore, it is appropriate to assume that S7450 and S7452 have the same intrinsic Si-Ge interdiffusivity. Based on the TEM measurements of TDD, the defect density of the top Ge layer, SiGe layer and the bottom Ge layer are close. Therefore, it is considered that for S7450, the Ge–Si interdiffusion only depends on the Ge content, which satisfies the condition of Boltzmann–Matano analysis.

Interdiffusivity values were extracted from the Ge SIMS profiles of S7450 using Boltzmann–Matano analysis which gives Ge–Si interdiffusion coefficient $\widetilde{D}(n_i)$ as a function of the Ge concentration at 750 °C and 800 °C respectively (Figure 7). Interdiffusivity data from two relevant studies were used for comparison [21, 25]. Dong et al. established a reference line for Si–Ge interdiffusivity with low TDD (about $10^5$ cm$^{-2}$) [21]. Gavelle et al. studied the interdiffusivity of highly defected Ge layer on Si substrate with a dislocation density of about $10^{10}$ cm$^{-2}$ [25]. The TDD of this work is on the order of $10^8$ cm$^{-2}$, which is in the middle. The acceleration effect of TDs is more obvious in the lower Ge region [25, 29]. For the high Ge region ($x_{Ge} > 0.9$), the extracted interdiffusivities of S7450 are within a factor of two compared to the reference diffusivities. Diffusivity difference of a few times is commonly seen in diffusion studies in semiconductors, which is mainly due to different materials quality (TDD levels), temperature calibration, and profile measurements. With this intrinsic interdiffusivity $\widetilde{D}(n_i)$, in the next section, we will build a model to describe the impact of the Fermi-level effect on Si-Ge interdiffusion.

## 4. Simulation Results and Predictions
4.1 Simulations of Extrinsic Si-Ge Interdiffusion

According to Equation (5), extrinsic interdiffusivity increases with the concentration of P. However, as shown in Figure 4 (a) that during annealing, P concentration profiles in S7452 changed due to P diffusion and segregation. Ideally, it is best to simulate P diffusion, P segregation and Si-Ge interdiffusion simultaneously. However, at this point, the diffusion and segregation coefficient of P on Ge concentration and the interdiffusivity dependence on P concentration are both unknown.

Therefore, in this study, an approximation method was used that treated P concentration profile as unchanged during the process of Si-Ge interdiffusion in S7452. This approximation is supported by two considerations: 1) the change from the as-grown P profile to the annealed P profile is small, which is from $2 \times 10^{19}$ to $8 \times 10^{18}$ cm$^{-3}$ in the most regions of interest, and 2) P diffuses much faster than Si-Ge interdiffusion, e.g., $D_P$ in SiGe is about 100 times larger than $\widetilde{D}_{SiGe}$ [21, 30, 31], it is assumed that P transportation reached the final profile much earlier than Si-Ge interdiffusion. Therefore, we took the annealed P concentration profile and selected the middle value in Figure 4 (a), i.e. P concentration profile in S7452 annealed at 800 °C for 30 minutes, as our background P doping profile in our simulation and assumed this profile was fast reached from the as-grown P profile and stayed unchanged during the interdiffusion process.

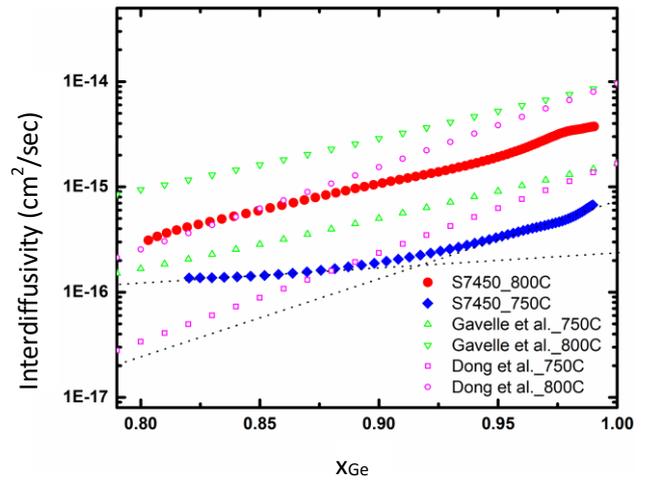

**Figure 7 Intrinsic Si-Ge interdiffusivity as a function of Ge molar fraction at 750 and 800 °C from S7450 using Boltzmann–Matano analyses in comparison with literature models at the same temperature. Dashed lines show the dislocation-mediated interdiffusivity and the point-defect-mediated interdiffusivity.**

The model fitting to data was done by Matlab using the models described in 3.1 and Fick's second law. To solve the diffusion equation numerically, finite difference time domain (FDTD) method was used. With different values of $m_1$ and $m_2$, we obtained several Ge profiles, as shown in Figure 8.

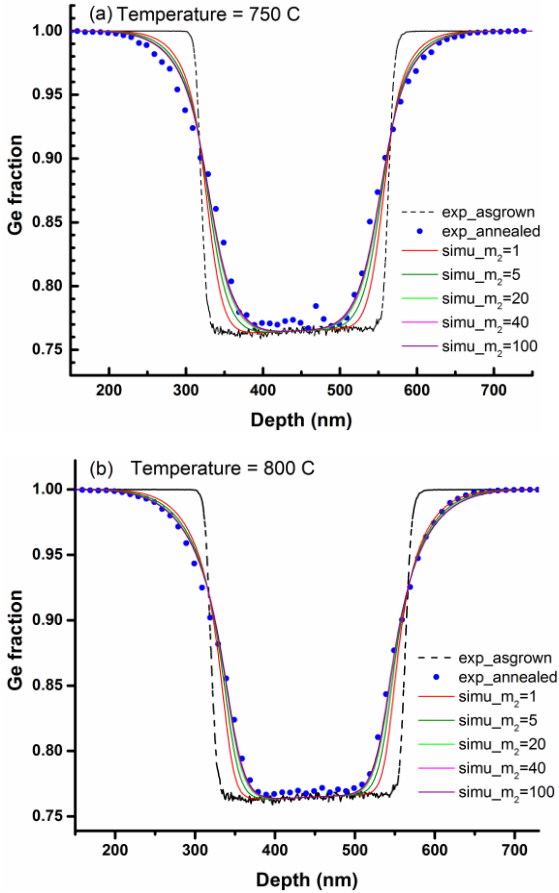

Figure 8 Comparison between SIMS data of S7452 and simulated results by using extrinsic Si-Ge interdiffusion model (a) At 750 °C for 120 minutes; (b) At 800 °C for 30 minutes. $m_1$ is fixed to 1 in each simulation and $m_2$ is 1, 5, 20, 40, 100 separately.

ure 8. The starting point was $m_1 = 1$ and $m_2 = 1$, and the simulated interdiffusion using these values at both temperatures was slower than the experimental interdiffusion profiles. Since increasing $m_1$ can at most increase FF slightly, according to Equation (12-16), $m_2$ was increased in our simulations to increase FF. As $m_2$ increases, interdiffusion becomes larger and closer to the experimental profiles. Moreover, the simulated results become insensitive with $m_2$ when $m_2 \geq 20$. This is consistent with Figure 7 (b), when $m_2 \geq 20$, FF increases very slowly with $m_2$ and is approaching $n^2/n_i^2$. For convenience, we chose $m_2 = 40$. Therefore, our simulation suggests that for $Si_{1-x}Ge_x$ with $0.75 < x_{Ge} < 1$, the quadratic term dominates, which means that the Si-Ge interdiffusion is dominated by $V^{2-}$ point defects.

With the best-fitting parameters ($m_1 = 1$, $m_2 = 40$), Figure 9 shows the value of FF in S7452. According to Equation (12), the impact of Fermi-level effect depends on the ratio of $n/n_i$ which is a function of P concentration. The FF profiles are noisy, which is resulted from the noisy P profiles from SIMS results. For low Ge fraction region (Ge valley in Figure 9), the interdiffusion has been enhanced by 3 to 7 times at 750 °C, and about 2-4 times at 800 °C.

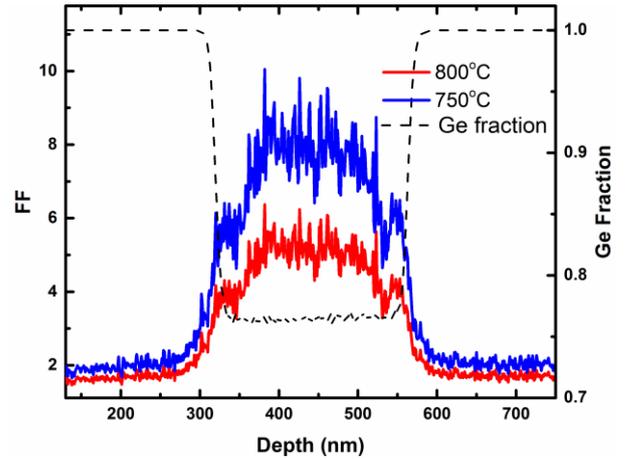

Figure 9 The values of FF at different temperatures as a function of depth. The dash line is the Ge profile of as-grown S7452. $m_1 = 1 \; and \; m_2 = 40$ were used in the calculation.

4.2 Model Predictions and Discussions

Finally, we used the models in Equation (12-16) and the best fitting parameters, $m_1 = 1$ and $m_2 = 40$, to predict FF as a function of $x_{Ge}$, n-type dopant concentration, and temperature. Figure 10 shows the dependence of Fermi-enhancement factor on different impacting factors. In Figure 10 (a), the FF is almost proportional to $n^2/n_i^2$. According to our simulation, for light n-type doping with $n/n_i = 2$, $x_{Ge} = 0.85$ at 750 °C, the FF is close to 4, which indicates the extrinsic interdiffusion coefficient is almost four times higher than intrinsic value. Therefore, in the case of heavy n-type doping, the Fermi-level effect on Si-Ge interdiffusion should not be ignored. In Figure 10 (b), for a SiGe wafer doping with P at $10^{19}$ cm$^{-3}$, as temperature decreases in the range of this study, the impact of the Fermi-level effect becomes more obvious. From 800 to 750°C, the enhancement of the FF is about 36%. In Figure 10 (c), for $n/n_i = 4$ at 750 °C, the FF increases with Ge fraction. However, our simulation shows this dependence is insignificant. The difference of FF between $x_{Ge} = 0.75$ and $x_{Ge} = 1$ is within 4%. This indicates that for $x_{Ge}$ in the range of 0.75- 1.00, the mechanism of Si-Ge interdiffusion is similar, i.e. $V^{2-}$ point defects are dominant for Si-Ge interdiffusion.

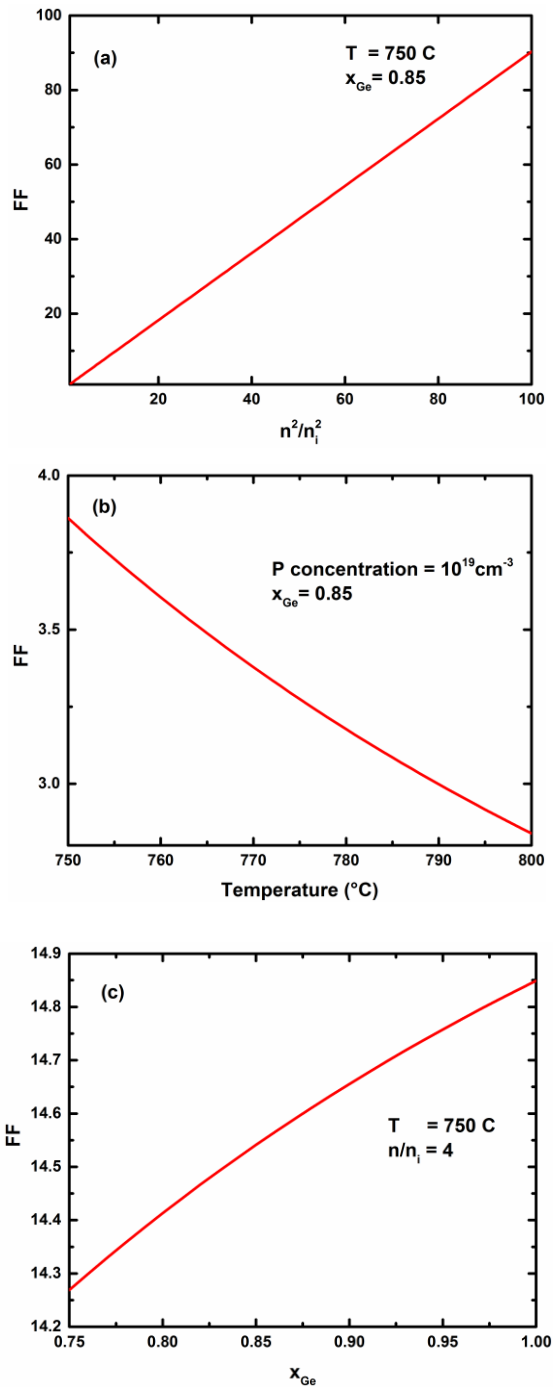

**Figure 10** Dependences of FF on different factors. a) FF dependence on $n/n_i$ at 750 °C and $x_{Ge} = 0.85$; (b) FF dependence on the temperature with P doping concentration of $10^{19}$ cm$^{-3}$ and $x_{Ge} = 0.85$; (c) FF dependence on $x_{Ge}$ at 750 °C and $n/n_i = 4$.

## 5. Conclusion

In summary, Si-Ge interdiffusion with a high P doping level was investigated by both experiments and modeling. Ge/Si$_{1-x}$Ge$_x$/Ge multi-layer structures with $0.75 < x_{Ge} < 1$, a mid-$10^{18}$ to low-$10^{19}$ cm$^{-3}$ P doping and a dislocation density of $10^8$ to $10^9$ cm$^{-2}$ range were studied. The P-doped sample shows an accelerated Si-Ge interdiffusivity, which is 2-8 times of that in the undoped sample. The doping dependence of the Si-Ge interdiffusion was modelled by a Fermi-enhancement factor. The results show that for Si-Ge interdiffusion coefficient is proportional to $n^2/n_i^2$ for the conditions studied, which indicates that the interdiffusion in high Ge fraction range with n-type doping is dominated by $V^{2-}$ defects. The Fermi-enhancement factor was shown to have a relatively weak dependence on the temperature and the Ge fraction. The results are relevant to structure and thermal processing condition design of n-type doped Ge/Si and Ge/SiGe based devices such as Ge/Si lasers.

**Acknowledgements**

This work was funded by Natural Science and Engineering Research Council of Canada (NSERC) and Crosslight Software Inc. The XRD measurements were performed in the Semiconductor Defect Spectroscopy Laboratory at Simon Fraser University. We are grateful for the help from Prof. Patricia Mooney on the XRD measurements. Dr. Stephen P. Smith at Evans Analytical Group is acknowledged for helpful discussions on SIMS measurements. Gary Riggott and Prof. Judy L. Hoyt from Microsystems Technology Laboratories, Massachusetts Institute of Technology are acknowledged for the epitaxy growth of the samples. Prof. Patricia Mooney from the Department of Physics at Simon Fraser University, Dr. Yiheng Lin and Guangnan Zhou from the Department of Materials Engineering, the University of British Columbia are acknowledged for the help in XRD measurements and helpful discussions.